\documentclass[12pt]{article}

\usepackage{amsmath}
\usepackage{amssymb,amsthm,amsxtra}

\usepackage{epsfig}   %% insert this in the preamble  now search
\usepackage{tikz-cd}
\newcommand{\ms}{\medskip}

\newcommand{\ba}{\begin{array}}
\newcommand{\ea}{\end{array}}

\newcounter{bean}
    {
      \begin{list}{\bf #1(\arabic{bean})}
         {\usecounter{bean}
              \setcounter{bean}{-1}
          \labelsep=1em
          \settowidth{\labelwidth}{#1\thebean:}
          \addtolength{\labelwidth}{1.1ex} 
          \leftmargin=\labelwidth 
          \addtolength{\leftmargin}{\labelsep} }

    }    {\end{list}}

\makeatletter
\@addtoreset{equation}{section}
\makeatother

\numberwithin{equation}{section}

\makeatletter
\def\iint{\DOTSI\protect\ints@\tw@}
\def\iiint{\DOTSI\protect\ints@\thr@@}
\def\iiiint{\DOTSI\protect\ints@{4}}
\def\idotsint{\DOTSI\protect\ints@\z@}

\def\intkern@{\mkern-6mu\mathchoice{\mkern-3mu}{}{}{}}
\let\DOTSI\relax
\let\ilimits@\displaylimits

\def\ints@#1{%
  \mkern-7mu\mathchoice{\mkern-2mu}{}{}{}%
  \mathop{\mkern7mu\mathchoice{\mkern2mu}{}{}{}%
    \intop\ifnum#1=\z@\intdots@
    \else\intkern@\fi
    \ifnum#1>\tw@\intop\intkern@\fi
    \ifnum#1>\thr@@\intop\intkern@\fi
    \intop
  }\ilimits@
}
\makeatother

\makeatletter
\def\@maketitle{\newpage
 \null
 \vskip 2em
 \begin{center}%
  {\normalsize\bf \@title \par}%
  \vskip 1.5em
  {\normalsize
   \lineskip .5em
   \begin{tabular}[t]{c}\@author
   \end{tabular}\par}%
  \vskip 2em
  {\@date}%
 \end{center}%
 \par
 \vskip 2.5em}
\makeatother
\makeatletter
\renewcommand\section{\@startsection {section}{1}{\z@}%
                                   {-3.5ex \@plus -1ex \@minus -.2ex}%
                                   {2.3ex \@plus.2ex}%
                                   {\normalfont\normalsize\bfseries}}
\renewcommand\subsection{\@startsection{subsection}{2}{\z@}%
                                     {-3.25ex\@plus -1ex \@minus -.2ex}%
                                     {1.5ex \@plus .2ex}%
                                     {\normalfont\normalsize\bfseries}}

\@addtoreset{equation}{section}
\makeatother

\begin{document}
\thispagestyle{empty}

\vspace*{3cm}

\begin{center}
\large{\bf   Directionality Theory and the Origin of Life  }

\vspace{1cm}
 \textbf{ Lloyd A. Demetrius}\\ 
  Dept. of Organismic and Evolutionary Biology,
  Harvard University\\ 
  Cambridge, Massachusetts 02138, U.S.A.\\
 March 12, 2023

\end{center}

\newpage
\begin{abstract}
	
The origin of cellular life can be described in terms  of the transition from  inorganic matter  --- solids, liquids and gases, to the emergence of  cooperative assemblies of organic matter,  DNA and proteins,capable of replication and metabolism.

\ms
Directionality Theory is a mathematical model of the collective behavior of populations of organic matter: cells and higher organisms. Evolutionary entropy, the cornerstone of the theory, is a statistical measure of the cooperativity of the interacting components that comprise the  population. The main tenet  of Directionality Theory is the \textit{Entropic Principle of Collective Behavior:} The collective behavior of aggregates of organic matter  is contingent on the population size and the external energy source, and characterized by extremal states of  evolutionary entropy.

\ms
This article invokes Directionality Theory to provide an evolutionary rationale for the following sequence of transformations which define the emergence of cellular life:
\begin{enumerate}
	\item The self--assembly of activated macromolecules from inorganic matter
	\item The emergence of an RNA world, defined by RNA molecules with catalytic and replicative properties
	\item The origin of cellular life, the integration of the three carbon--based polymers --- DNA, proteins and lipids, to generate a metabolic and replicative unit.
\end{enumerate}

\end{abstract}	
	\newpage

\centerline{\large\bf{1.  Introduction}}

\ms
Cellular life, the fundamental unit of living organisms, consists of an elaborate set of self--reproducing chemical reactions. These reactions rely upon a large set of specific catalysts, encoded by a DNA genome. The metabolic and replicative components of cells are determined by  three classes of carbon--based polymers:
\begin{enumerate}
	\item  \textit{Amphiphiles: } These polymers  form the boundary membranes of all cellular life
	\item  \textit{Proteins: } These biopolymers are synthesized by linking together amino acids using an external source of energy. A group of proteins, called enzymes, have the unique ability to act as catalysts 
	\item  \textit{DNA: } The subunits of these polymers are nucleotides. DNA stores and transmits genetic information.
\end{enumerate}	
Cellular life is a process in which the genetic and catalytic polymers coordinate a dynamical system where  information in the DNA directs the synthesis of  proteins; and the catalytic activity of proteins  regulates the synthesis of  DNA.

\ms
Modern cellular systems are isothermal, chemical machines that appropriate energy from the external environment, and convert this energy into metabolic energy and biological work. During growth, the cyclic system of polymers reproduces itself. Random mutations in the DNA entail that differences in the metabolic capacity of the daughter cells will occur. Since, different cells vary in their capacity to appropriate an external energy source  into biomass, selection will occur, and the composition of the population will evolve.

\ms
These observations suggest that the problem regarding the origin of life   can be articulated as follows: 
\textit{What are the physico--chemical  principles which underlie the transition from the abiotic molecules that define condensed matter --- solids, liquids and gases, to the integrated assembly of organic matter  that characterizes cellular life?}

\ms
The empirical studies began by Harold Urey and Stanley Miller in 1953, have provided a conceptual template for addressing  the problem. The Miller--Urey studies showed that organic molecules  could be created from inorganic matter by environmental conditions such as heat and  electric discharge, without the mediation of enzymes. This discovery suggested that the transition from inorganic matter to  activated biomolecules,  the basic components of  organic matter, can occur spontaneously, provided there exist an external source of energy.

\ms
However,  the transition from the state of activated biomolecules to cellular life appeared intractable,  due largely to the complexity of the processes which appeared to underlie the origin of proteins and DNA.

\ms
Empirical studies of the catalysis of metabolic reactions in modern cells have elucidated the transition from activated biomolecules to cellular life by delineating  two  kinds of catalytic mechanisms, Alberts (1986), namely:
\begin{enumerate}
	\item[(a)]  \textit{Class I Reactions: } These metabolic processes are catalyzed by ribonucleoprotein complexes. These reactions are RNA--based, and first evolved in cells that contained no protein or DNA.
	\item[(b)]  \textit{Class II Reactions: } The  molecules which are involved in these systems are proteins with an elaborately--folded catalytic surface. These molecules emerged relatively late in the evolutionary history of the cell.	
\end{enumerate}		
The discovery, Cech (1985), that RNA  can act as both a catalyst and a template, suggested that modern--day cells are the evolutionary derivatives of an RNA world in which all molecular catalysts were ribozymes.

\ms
RNA is a less effective information and catalytic polymer  than the DNA and protein that perform these functions in modern cells. DNA is more stable and less susceptible to mutation than RNA. Proteins have more complex three dimensional structures than RNA. 
These characteristic features of  RNA  have provided  an evolutionary rationale for the existence of an RNA world, and its subsequent replacement by a system whose operational units are DNA and Protein, Alberts (1986), Copley et al. (2007), Higgs and Lehman (2015). 

\ms
Accordingly, the transformation from the emergence of activated bio--molecules  --- carbon--based monomers, to the organization of cellular life can be understood in terms of the following transitions,  Alberts (1986), Demongeot and Thellier (2023), Demongeot and Seligmann (2022).
\begin{enumerate}
	\item[(a)]   \textit{The emergence of RNA as an informational polymer}\\
	The support for this  emergence derives from the capacity of RNA to perform with the help of metals, and other small molecules, reactions which   exhibit template  and catalytic activity.
\item[(b)]  \textit{The transition from RNA to protein: }\\
RNA can fold in complex ways that endow the molecules with the catalytic activity associated with proteins. Proteins, however, are more effective catalysts.
\item[(c)]  \textit{The transition from RNA to DNA: }\\ 
RNA can also fold to assume the double helix structure which enables storage of genetic information. DNA, however,  is more stable and less vulnerable to replication errors.
\end{enumerate}
The pioneering investigations of Urey and Miller (1953), and the recent discoveries regarding the RNA world, as reviewed in Alberts (1986), Copley et al. (2007), have stimulated a large body of  theoretical  and empirical research concerning the emergence of cellular  life.  Recent theoretical contributions, for example  Dyson (2004), Eigen (1992), Pross (2012), Walker (2017), and Davies (1998) have focused on  questions regarding replication or metabolism as initiating the emergence of cellular life. 

\ms
Several theoretical contributions have implicitly rejected the existence of an RNA world, and the subsequent emergence of proteins as the more effective catalyst, and DNA as the more stable information carrier.  Christian de Duve (1995) replaced the RNA world by a model which describes  how a proto--metabolic web may have emerged prior to the appearance of membranes and of cells. The computational models of  Kauffman (1993) is based on the interaction of prebiotic precursors  to generate autocatalytic networks. 

\ms
Hartman (1989)  also rejects the hypothesis of an RNA world, and postulates the origin and evolution of photosynthesis as the cardinal factor which underlies the transition to organic matter and cellular life.

\ms
Empirical studies of the origin of life have been rooted in the biochemical implications of the RNA world as the standard model. Experiments, as described in Szostak et al. 2001), confirm,  in large part that the catalytic capabilities of RNA are consistent with the RNA world hypothesis. Deamer (2011) provides an extensive critical evaluation of the biochemical and bioenergetic implications of the RNA world and its pertinence in elucidating the origin of cellular life. The investigations reported in  Deamer and Szostak (2010), succeeds in placing the RNA world hypothesis on firm ground through a set of experimental studies of the catalytic and templating features of RNA.

\ms
The mathematical models which have been proposed to investigate  the emergence of life have been largely influenced by the fact that collective behavior, that is the  sharing and distribution of energy among the components of a system, is one of the hallmarks of living organisms. 

\ms
There exist two analytical models  of collective behavior --- Statistical Thermodynamics and Directionality Theory. These two methodologies have  emerged as competing conceptual templates in analytic studies of the origin of life. 

\ms
Statistical Thermodynamics is concerned with deducing the Thermodynamic properties of a macroscopic system from the cooperative interaction of its microscopic components. These components are the atoms and molecules which comprise inanimate matter --- solids, liquids and gases. 

\ms
Directionality Theory is concerned with elucidating the adaptive properties of a macroscopic network  from the cooperative interaction of the individual components. These components are the cells that comprise a tissue, or the higher organisms that define a population. Statistical Thermodynamics and Directionality Theory are both based on the fundamental idea that the parameter which encodes the collective behavior of a macroscopic system  are statistical measures of the cooperativity that defines the interaction between the components. 

\ms
The index of cooperativity invoked in Statistical Thermodynamics is Thermodynamic entropy, a statistical measure of the number of instantaneous microstates consistent with a given macrostate.

\ms
Cooperativity in Directionality Theory is encoded by Evolutionary entropy. This statistical parameter  incorporates the statistical dependence and the temporal progression of microstates which arise from the long--range interaction between the components of the  macroscopic system. These two measures of cooperativity, Thermodynamic entropy and Evolutionary entropy, are related are related.

\ms
Thermodynamic entropy is the limit, as $N\to \infty$ (where $N$ denotes the number of degrees of freedom  of the system) of Evolutionary entropy.

\ms
The basic premise of Statistical Thermodynamics is expressed in terms of a directionality principle which describes the collective behavior of the atoms and molecules in a macroscopic aggregate. The analysis of the changes in thermodynamic entropy due to collision between the molecules in the aggregate, assumes that the system is isolated, and closed to the input of energy and matter. The sharing and spreading of energy among the molecules of the aggregate is expressed by the following principle.

\ms
\textit{The Second Law of thermodynamics: } The collective behavior of the interacting molecules is characterized by the maximization of thermodynamic entropy.

\ms
The cornerstone of Directionality theory is a set of principles which encode changes in evolutionary entropy due to variation, a process by which new types are introduced in the population,  and selection, which discriminates between the variant types according to their capacity to convert external sources of energy to biological and physico--chemical  work.

\ms
The principles which describe changes in evolutionary entropy will be contingent on the mechanism, genetic mutation, or macromolecular  fluctuation, which generates the variability that characterizes  the components. 

\ms
Variation in evolutionary  systems,  the collective behavior of replicating and metabolizing entities,  is determined by mutation in DNA, the biochemical support for the storage of genetic information. The concomitant changes in evolutionary entropy, due to combined effect of  mutation and natural selection, is encoded by the following rule:

\ms
\textit{The Entropic Principle of Evolution: } The collective behavior of an evolutionary  system is contingent on the population size, the external resource endowment, and characterized by extremal states of evolutionary entropy.

\ms

Variation in Self--organizing processes, for example the collective behavior of organic molecules, effecting molecular recognition, is described by fluctuations. This generic term refers to changes in the constitution of the macromolecules due to the  lability of the interactions connecting the components. The resulting changes in evolutionary entropy due to fluctuation and natural selection, is formalized in terms of the following tenet:

\ms
\textit{The Entropic Principle of Self--Organization:} The collective behavior of a self--organizing system  is contingent on the dynamics of the energy source and characterized by states which maximize evolutionary entropy.

\ms
This article will invoke Directionality Theory, to propose a physico--chemical mechanism to characterize the transition from inanimate matter to cellular life. We will show that the transition from inorganic matter to protocells can be explained in terms of the \textit{Entropic Principle of Self--Organization.}

\ms
Protocells are compartmented systems composed of two types of RNA polymers, namely the information units, the precursors to DNA, and the  catalytic units, the precursors to proteins. Experimental studies in support of the emergence of protocells from lipid extracts and RNA polymers have been described in Deamer (1997).

\ms
We will  then show that the transition from protocells to cellular life can be resolved in terms of the Entropic Principle of Evolution.

\ms
This article is organized as follows. Section 2 will contrast the two methodologies, Statistical Thermodynamics and Directionality Theory, with regards to their pertinence  as conceptual templates for elucidating the emergence of cellular life from inorganic matter.

\ms
Section 3 presents a qualitative description of evolutionary entropy and the methodology of Directionality Theory. Section 4 is an application of  Directionality Theory to explain the emergence of life as a two--step process:
\begin{enumerate}
	\item[(a)] The transition from inanimate matter to protocells --- a process driven by Self--organization
\item[(b)] The transition from protocells to cellular life --- a process realized by Darwinian evolution.
\end{enumerate}

\newpage
{\large\bf{2.  Collective behavior: Statistical Thermodynamics and Directionality Theory}}

\smallskip
The  physico--chemical principles which underlie the transition from abiotic molecules to the integrated assembly of organic matter encodes the  collective behavior of the macroscopic system, and the cooperativity of the interacting components, inorganic and organic.

\bigskip
\textbf{2.1. Collective behavior: Statistical Thermodynamics}

\smallskip
Equilibrium statistical thermodynamics constitutes a very powerful theory for explaining   collective behavior in aggregates of inanimate matter. A cornerstone of the theory is the concept Thermodynamic entropy, which Boltzmann introduced as a statistical measure of the cooperative interaction of the atoms and molecules that comprise a macroscopic system. 

\ms
Boltzmann's study of the relation between the thermodynamic properties of a gas and the collective behavior of the molecules is expressed in terms of the Second Law of Thermodynamics: \textit{Thermodynamic entropy increases in systems which are isolated and closed to the input of energy and matter.}

\ms
The principle however, does not explain the collective behavior observed in populations of abiotic molecules or cellular populations.  Recent extensions of equilibrium statistical thermodynamics to investigate collective  behavior in  biological systems  are based on a class of propositions, called Fluctuation Theorems, Prigogine and Nicolis (1977), Crooks (1999), Jarzynski (2011).

\ms
Fluctuation theorems constitute the cornerstone of non--equilibrium statistical thermodynamics. The application of this methodology to the study of the origin of cellular life is largely due to the pioneering research of Prigogine and Nicolis (1971). 

\ms
Michaelian (2011), England (2013), Davies et al. (2017),  Endries (2017), Andrien and Gaspard (2008)  are recent efforts to apply the thermodynamic formalism to study the transition from inorganic matter to cellular life. These studies have made evident the conceptual problems which arise in investigating the origin of life from the perspective of statistical thermodynamics.

\ms
The failure of the thermodynamic methodology to elucidate   the emergence of cellular life derives from two main considerations:
\begin{enumerate}
	\item[(i)] Thermodynamic entropy is not a valid  measure of cooperativity  for the collective behavior of aggregates of organic matter.
\end{enumerate}
Cooperativity between the microscopic components in organic matter involve long range interactions and statistical dependencies  between the microscopic components. Thermodynamic   entropy, a valid measure of cooperativity in aggregates of inorganic matter, does not incorporate the statistical dependencies and the  temporal progression of microstates which is a characteristic feature of the collective behavior of organic matter.
\begin{enumerate}
	\item[(ii)] Heat is not a viable currency for the transfer of information and energy  in biological systems.
\end{enumerate}
Energy, the capacity to do work is expressed in two major forms, potential and kinetic. Potential energy is energy associated with the position, configuration or shape of an object in a force field. Kinetic energy is the energy which is associated with a body due to the motion of its component elements.

\ms
The forms of energy which drive biological process are all potential. These sources include: 
\begin{enumerate}
	\item[(i)] The chemical energy which is contained in the electron bonds that hold atoms together.
\item[(ii)] The redox--potential which is manifest during the motion of electrons.
\item[(iii)] The photochemical energy which emerges when light is absorbed by the electron structure of a pigment molecule.
\item[(iv)] The chemiosmotic energy induced in concentration gradients across  membranes.
\end{enumerate}
Thermal energy is random kinetic energy. The transfer of heat is contingent on the difference in temperature between the source and the sink. The maximum work which may be derived in heat transfer is given by the expression
$$
w=q\left(\frac{T_2-T_1}{T_2}\right)
$$
The quantity $q$ is the heat absorbed, and $T_2$ and $T_1$  are the absolute temperature of the source and sink. 

\ms
The temperature differences in different parts of a cell are insignificant. Hence, $\omega\sim 0$. This condition entails that thermal energy is not a useful way of effecting the biological work, synthetic, osmotic and muscular, which is necessary to maintain the stability of the living state. Cells are unable to function as heat engines.

\ms
These considerations, the limitations of thermodynamic entropy as a measure of cooperativity, and the isothermal nature of energy transformation in biological systems, are the two main factors that invalidate the various efforts to apply  extensions of the thermodynamic formalism of  Onsager (1931) and Prigogine (1969) to the study of collective behavior in biological systems.

\pagebreak
\textbf{2.2. Collective behavior and Directionality Theory
}

\smallskip
Directionality theory is a mathematical model of  collective behavior in aggregates of organic matter: activated macromolecules, cells, higher organisms. The theory was originally developed to study  the  evolution of populations where the state of individuals is parametrized by age, Demetrius (1974), (1975).  The concept evolutionary entropy was introduced as a measure of cooperativity in these demographic models. Semelparous populations, such as annual plants, have zero entropy. Reproduction in these populations occurs at a single stage in the life--cycle. Perennials, organisms whose reproductive activity is distributed over several age--classes, have positive evolutionary entropy. 

\ms
Later studies, Demetrius (1983), have generalized the concept  of evolutionary entropy to describe cooperativity in populations where the state of individuals is parametrized by morphological, physiological or behavioral properties. 
Evolutionary entropy has now emerged as a generic measure of cooperativity, a parameter which describes the statistical dependencies and correlations induced by  the interactions between individuals defined in terms of their microstates.

\ms
Directionality Theory is the study of dynamical changes in evolutionary entropy under the processes of variation and natural selection. Evolutionary dynamics is a study of the dynamics of collective behavior when the variation process  is due to mutation. The changes in evolutionary entropy under this constraint is described by the following tenet:

\ms
\textit{The Entropic Principle  of Evolution:}  The collective behavior  of metabolic, replicative entities, subject to the evolutionary process of mutation and natural selection is  contingent on the population size  and the external energy source.The steady states of the evolutionary process are extremal states of evolutionary entropy.

\ms
Self--organization, the spontaneous emergence of spatio--temporal order due to local interactions between the metabolic entities is an example of collective behavior where the variation among the components is induced by the lability of the interaction connecting the individual units. The changes in cooperativity is now determined by a fluctuation--selection process. These changes, a derivative of  Directionality Theory are described by the following tenet:

\ms
\textit{The Entropic Principle of Self--Organization:} The collective behavior of organic matter subject to the self--organizing process of fluctuation and selection, is contingent on the external energy source and independent of the population size. The steady state of the self--organizing process are states which maximize evolutionary entropy.

\ms
Our analysis of the emergence of cellular life will distinguish between two phases
\begin{enumerate}
	\item[(I)] The transition from inorganic matter to protocells, defined by an assembly of RNA polymers.
 	\item[(II)]	The transition from protocells to cellular life, as defined by the integrated assembly of DNA, proteins, RNA and lipids.
	\end{enumerate}
We will show that Phase (I) can be analyzed in terms of Self--assembly. Variation in processes of self--assembly is induced by fluctuations --- changes generated by the lability of the interaction connecting the individual units. Accordingly, Phase (I) can be explained in terms of the \textit{Entropic Principle of Self--Organization.} 

\ms
We will show furthermore that Phase (II) falls within the Darwinian paradigm. Variation is induced to mutation in the information polymers. The transition in this case will be analyzed in terms of the \textit{Entropic Principle of Evolution.}

\newpage
\textbf{\large{3. Evolutionary Entropy and Collective Behavior}}

\smallskip
Collective behavior is a generic term which characterizes the evolutionary dynamics of a population of interacting components  and their adaptation to a given environment. We refer to Demetrius (2013) for a detailed exposition of the model we now describe.

\ms
A population, the fundamental unit in models of collective behavior, is described as a strongly connected, directed graph, as shown in Fig, 1.

\ms
\begin{center}
	\includegraphics[width=6cm]{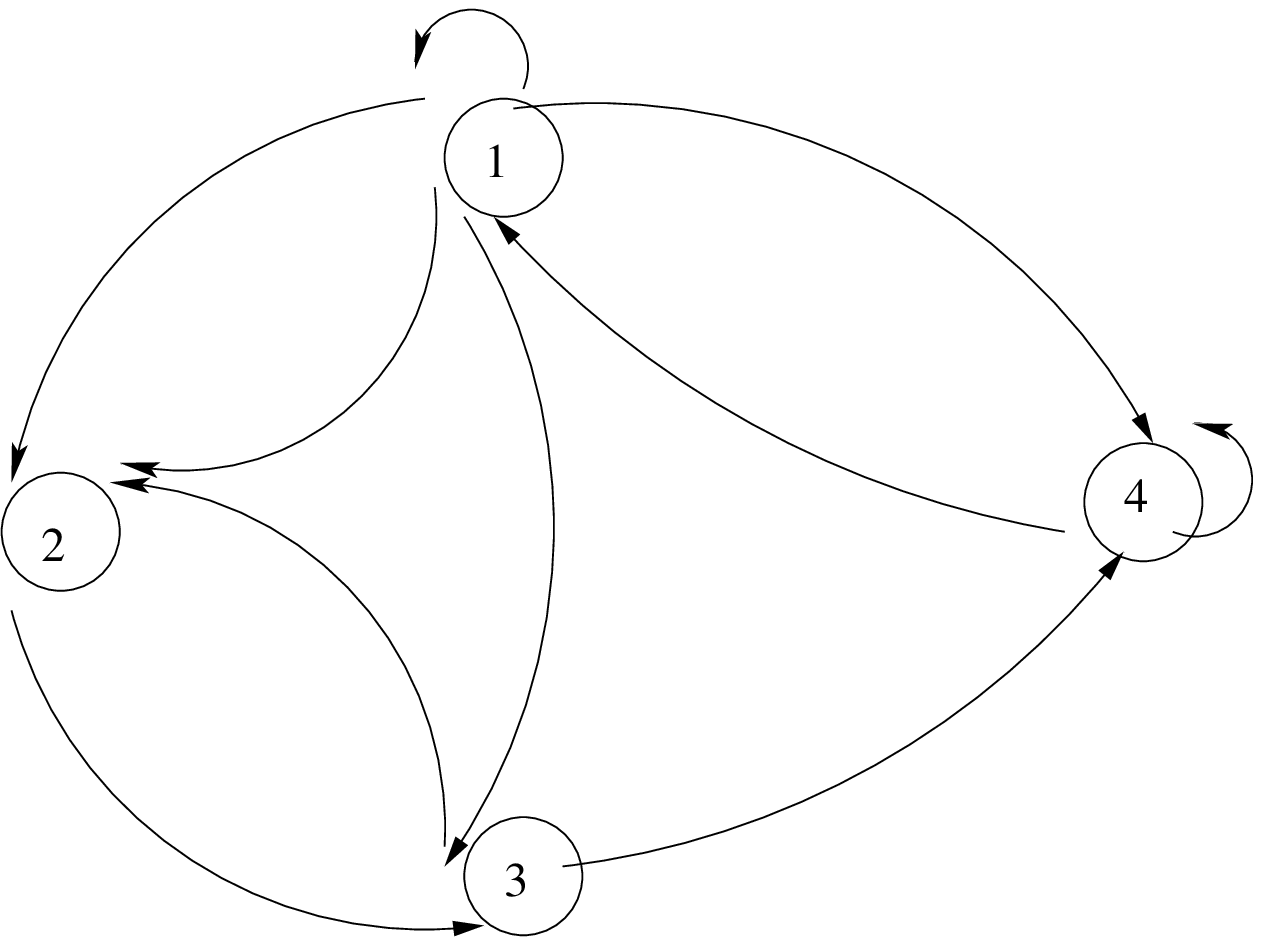}

\smallskip
Fig. 1: Population as a strongly, connected, directed  graph
\end{center}	
	\ms
The nodes of the graph represent the microstates of the system, that is the set of physical or chemical properties that describe the components. The links between the nodes describe the interaction between the components.

\ms
We denote the set of nodes in the graph by the set  $S=(1,2,\ldots d)$. The model is described by the mathematical object $(\Omega,\mu,\varphi)$, whose  elements are defined as follows:
\begin{enumerate}
	\item[(i)] $\Omega$: The set of all paths or sequences generated by the graph. A path is a sequence
	$$
	x=(\ldots x_{-1}x_0x_1\ldots)
	$$
	where $x_i$ is an element in the set $S$.
	\item[(ii)]  $\varphi$: a rule which associates  with each element $x\in \Omega$, a number which describes the interaction between the components that generate  the sequence.
	\item[(iii)] $\mu$: a probability measure which describes the frequency distribution of the different sequence that define the network. The parameter $\mu$ characterizes the steady state value  of the dynamical system induced by the interaction .
\end{enumerate}	
The statistical parameter, evolutionary entropy, denoted $H$, is defined by $H=S/T$, where
$$
S=-\sum_{\tilde{\alpha}\in C_\alpha}p_{\tilde{\alpha}}\log p_{\tilde{\alpha}}\, ; \quad T=\sum_{\tilde{\alpha}\in C_\alpha}|\tilde{\alpha}|p_{\tilde{\alpha}}\leqno(1)
$$
The element $\alpha$ denote a microstate, whereas $C_\alpha$ denote the set of trajectories that begin with $\alpha$, ends  at $\alpha$ without traversing $\alpha$ as an intermediate state. $C_\alpha$ is the temporal progression of instantaneous states which begin at the mode $\alpha$. For a given element ${\tilde{\alpha}} \in C_\alpha$ the quantity $p_{\tilde{\alpha}}$ is the probability of the cycle defined by the trajectory $\tilde{\alpha}$.

\ms
The evolutionary entropy $H$, which has its roots in the ergodic theory of dynamical systems, Demetrius (1974), (1987), is a measure of Cooperativity of the interacting components that define the network. Cooperativity describes  the extent to which the components are held together by the forces which effect the interaction between the components. 

\ms
The model assumes that the population appropriates resources or energy from an external environment, and converts this energy into metabolic energy and biological work. The conversion of the resource $R(t)$ into population number $N(t)$ is represented in Fig. 2

\ms
\begin{center}
	\includegraphics[width=12cm]{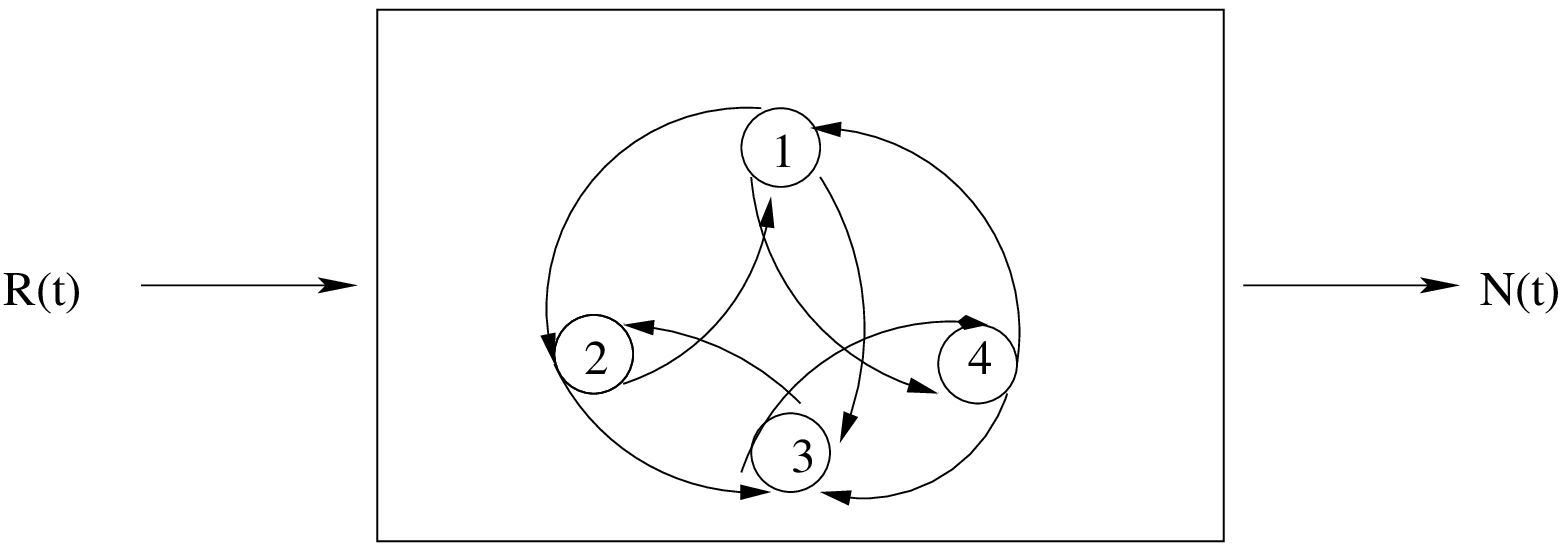}

\smallskip
Fig. 2: Conversion of the resource endowment $R(t)$ into population numbers $N(t)$
\end{center}

\ms
The demographic parameters that describe the changes in population numbers induced by the  conversion of the resource endowment are the population growth rate $r$, and the demographic variance $\sigma^2$.

\ms
The resource parameters  for the growth and adaptation of the population are the energy production rate $\Phi$, and the temporal correlation of the energy production $\gamma$. These parameters are correlated with the demographic parameters $r$ and $\sigma^2$ which describe the population dynamics. We have
$$
\Phi=\frac{dr(\delta)}{d\delta}\bigg|_{\delta=0}\, \, ; \, \, \gamma =  \frac{d\sigma^2(\delta)}{d\delta}\bigg|_{\delta=0}\leqno(2)
$$
The quantities $r(\delta)$ and $\sigma^2(\delta)$ are the growth rate and the demographic variance of the variant population $(\Omega,\mu(\delta),\varphi(\delta))$.

\ms
The function $\varphi(\delta)$ is given by
$$
\varphi(\delta)=\varphi+\delta\varphi$$
where $\delta$ is a small perturbation.  $(\Omega,\mu,\varphi)$ is assumed to be the resident population of size $N$, and evolutionary entropy $H$. 

\ms
 $(\Omega,\mu(\delta),\varphi(\delta))$ is assumed to be  a variant population of size $N^*$, and evolutionary entropy $H^*$.

\ms
Directionality Theory is the study of the outcome of competition between the incumbent and the variant population, assuming that $N^* \ll N$.

\ms
Let
$$
\Delta H = H^*-H
$$
The theory is concerned with directional changes  in evolutionary entropy  as a result of the outcome of competition between the incumbent population, $(\Omega, \mu(\delta),\varphi(\delta))$. 

\ms
The rules which describe the changes  in evolutionary entropy, will be contingent on the variation process and the external energy source. We will illustrate the effect by considering the dynamics of collective behavior  in systems where the individual components are cells and macromolecules, respectively.

\bigskip
\textbf{3.1 Cellular systems: Mutation--Selection Process}

\smallskip
Cellular systems are organized in terms of a genetic  and a  metabolic component. Variation in cellular systems is generated by  mutations,  that is changes in the information carrier, DNA. Mutations will induce changes in the metabolic component,  and thereby generate alterations in cellular behavior. 

\ms
The dynamics of collective behavior  at the cellular level  implicates three processes: Mutation --- random changes in DNA; Inheritance --- the transfer of the information from the mother cell to the daughter cells.   Selection --- competition between the variant  and incumbent types for the resource endowment.

\ms
Mutation induces an \textit{irreversible }process. If the resident population is described by the demographic parameters $(r,\sigma^2)$, and the variant population by the parameter $(r(\delta),\sigma^2(\delta)$, then $\Delta r \neq 0,\, \, \Delta\sigma^2\neq 0$.

\ms
The resource constraints $\Phi$ and $\gamma$, as defined by (2), will   therefore satisfy the condition $\Phi\neq 0$, $\gamma \neq 0$.

\ms
We showed, Demetrius (2013), that the rules which characterize the change in evolutionary entropy  due to mutation and selection are expressed by the following tenet:

\ms
\textit{The Entropic Principle of   Evolution:}  The outcome of variation and natural selection on collective behavior is contingent on the population size, the external energy source and characterized by extremal states of evolutionary entropy.

\ms
The analytic description of the principle is the relation
$$
(-\Phi+\gamma/M)\Delta H \ge 0\leqno(3)
$$
Here $M$ denote the total population size.

\ms
The \textit{Entropic  Principle of Evolution} pertains to Darwinian systems, that is evolutionary change by mutation and natural selection. The principle is applicable to evolution at both cellular and multi--cellular levels of organization. 

\ms
The relation between the statistical parameters $\Phi, \gamma$ and the directional change $\Delta H$, is summarized in Table (1) and Table (2)

\ms
\begin{center}
	\textbf{Table (1)} ($\Phi\gamma < 0$): The relation between the statistical parameters $\Phi\gamma$ and the directional changes in evolutionary entropy
	
	\ms
	\begin{tabular}{|l|l|}
		\hline
		\textbf{Resource constraints}&\textbf{Directional Change in Evolutionary Entropy}\\
		$\Phi < 0,\, \, \gamma > 0$&$\Delta H > 0$\\
		$\Phi > 0,\, \, \gamma < 0$&$\Delta H < 0$\\
		\hline
	\end{tabular}
	
	\vspace{1cm}
	\textbf{Table (2)} ($\Phi\gamma > 0$): The relation between the statistical parameters $\Phi\gamma$,  the population size $M$, and directional changes in evolutionary entropy
	
	\ms
	\begin{tabular}{|l|l|}
		\hline
		\textbf{Resource constraints and}&\textbf{Directional Change in Evolutionary Entropy}\\
		\textbf{ Population size}& \\
		\framebox{$\Phi < 0, \, \, \gamma < 0$}& \\
		\quad $\gamma < M\Phi$&$\Delta H > 0$\\
		\quad $\gamma > M\Phi$&$\Delta H < 0$\\
	\framebox{$\Phi > 0,\, \, \gamma > 0$}& \\
		\quad $\gamma < M\Phi$&$\Delta H < 0$\\
		\quad $\gamma > M\Phi$&$\Delta H > 0$\\       
		\hline
	\end{tabular}
\end{center}

\ms
Empirical support for the Entropic Principle of Evolution  in demographic and metabolic networks is described in Demetrius (2013).

\bigskip
\textbf{3.2 Molecular Systems: Fluctuation--Selection Process}

\smallskip
Molecular systems are  macromolecular  structures, whose components are held together by non--covalent intermolecular forces. Variation at molecular scales  derive from the lability of the interaction connecting the components. Molecular instability entails  that chemical  bonds may form and break reversibly, thus leading  to a continuous  reorganization  of the components. The dynamics of collective behavior will now implicate two processes: Fluctuation and selection.

\ms
{Fluctuation} is a generic term which  describes the changes in structure induced by the lability of the molecular interactions. 
 Selection involves  competition between the variant  components for the external chemical  energy.

\ms
Fluctuation is a \textit{reversible} process. The resident and variant populations  are described by the demographic parameter $(r,\sigma^2)$, and $(r(\delta),\sigma^2(\delta))$ respectively.

\ms
Since the variation process is driven by the lability of the molecular interaction, $\Delta r \neq 0$, and $\Delta\sigma^2=0$. We conclude from (2), that $\Phi\neq 0$ and $\gamma = 0$.

\ms
Consequently, the changes in evolutionary entropy induced by the fluctuation--selection process will only implicate $\Phi$, and be independent of $\gamma$, and the population size.  These changes are formalized by the following principle:

\ms
\textit{The Entropic Principle of Self--Organization:} The outcome of variation and selection on the collective behavior of the system is contingent on the energy constraint and characterized by the maximization of evolutionary entropy.

\ms
The entropic principle of Self--organization is given by the relation
$$
-\Phi\Delta H > 0\leqno(4)
$$
Table (3) relates the resource constraints with directional changes in evolutionary entropy

\ms
\begin{center}
	\textbf{Table (3)} Directional Changes in Evolutionary Entropy
	
	\ms
	\begin{tabular}{|l|l|}
		\hline
		\textbf{Resource constraints}&\textbf{Directional Change in Evolutionary Entropy}\\
		$\Phi < 0$&$\Delta H > 0$\\
		$\Phi > 0$&$\Delta H < 0$\\
		\hline
	\end{tabular}
\end{center}

\bigskip
\textbf{3.3 Static and Dynamic Self--Assembly}

\smallskip
The relation given by (3) is a general rule for the emergence of Self--organization. This rule pertains to  two mechanisms of self--assembly: static and dynamic. The distinction between these two modes is based on the following identity
$$
r=H+\Phi\leqno(5)
$$
This relates the population growth rate $r$, with the evolutionary entropy $H$, and the reproductive potential $\Phi$.

\ms
Static self--assembly describes processes in which there is no dissipation of energy. These systems satisfy the relation $r=0$. 

\ms
Static self--assembly entails the relation
$$
\Delta H \ge 0\leqno(6)
$$
We can infer the following rule:

\ms
\textit{Entropic Principle of Static Self--Assembly: } The equilibrium state for the process of static self--assembly are states which maximize evolutionary entropy.

\ms
Dynamic self--assembly describes dissipative processes. These processes are characterized by the condition $r\neq 0$.

\ms
We can appeal to the relation between (4) and (5), to show that dynamic self--assembly implies the relation
$$
(H-r)\Delta H \ge 0\leqno(7)
$$
The condition (7) yields the following qualitative statement:

\ms
\textit{The Entropic Principle of Dynamic Self--Assembly:}  The equilibrium state of the process of dynamic self--assembly is a steady state which maximizes evolutionary entropy, contingent on the relation between evolutionary entropy and the population growth rate.

\ms
The article will appeal to the\textit{ Entropic Principle of Evolution, } as described by (3) to propose an evolutionary rationale for the transition, proto--cells to cellular life. 

\ms
The \textit{Entropic Principle of Self--Organization,} the tenet described by (4), will be invoked to propose a mechanistic model to explain the transition from inorganic matter to protocells.

\newpage
\textbf{\large 4. The Emergence of Cellular Life}

\smallskip
The transformation from inorganic matter to the emergence of chemical assemblies capable of  evolution by mutation and natural selection can be depicted in terms of the following transitions. 

\ms
\textbf{(I) } \textit{The transition from Inorganic matter to Protocells	}

\smallskip
This process involves the  transformation of inorganic matter to a class of carbon--based activated macromolecules, amphiphiles, nucleic acids. 

\ms
The emergence of the  protocellular state can be depicted in terms of the following phases:

\ms
(a) \textit{Amphiphiles to vesicular polymers:}\\
The polymerization of amphiphilic molecules to form stable vesicles 

\ms
(b) \textit{Nucleic acids to Informational Polymers: The RNA world\\}
The polymerization of nucleic acids to generate  RNA with catalytic and informational capabilities. 

(c) \textit{The emergence of protocells\\}
The transformation to the protocellular state.  Protocells are characterized by a membrane that defines a spatially localized compartment, and an informational polymer that enables the replication and inheritance of functional information.

\bigskip
\textbf{(II) }\textit{ The transition from protocells to cellular life}

\smallskip
The emergence of chemical assemblies defined in terms of the four carbon--based polymers, lipids, DNA, RNA, protein. Empirical and experimental support for the transition depicted by (I) is reported in various studies, Chen  et al. (2004), Szostak et al. (2001), Deamer (2011).

\ms  
 We will show that each of the phases in (I) ---  the transition  from inorganic  matter to protocells --- can be explained in terms of Self--assembly processes.  This mechanism is formalized in terms of the \textit{Entropic Principle of Evolution.}
 
 \ms
 We will then show that phase (II),  the  transition from protocells to cellular life, can be formalized in terms of the \textit{Entropic Principle  of Selection}

\ms
\textbf{4.1 The transition from Inorganic matter to protocells.}

\smallskip
Geological and geophysical evidence indicates that the earth's atmosphere was once reducing and composed predominantly of the inorganic states of matter: methane, nitrogen and ammoniac.

\ms
A protocell,  a fundamental unit of life, is an assembly of a class of carbon--based activated macromolecules;   defined by  a lipid membrane, and RNA polymers with replicative and catalytic function.

\ms
The membrane forms an enclosed compartment which provides the protocell with aqueous spaces. The metabolic and replicative components function together and form an autocatalytic system. Protocells are the primitive precursors to cellular life defined in terms of the three carbon--based polymers, lipids, DNA and proteins. The transition from an aggregate of inorganic matter to a protocell can be described in terms of the following sequence of transitions

\ms
\textbf{(I) Inorganic matter to Activated Biomolecules :} (\textit{Dynamic Self--Assembly})
\\
Experimental studies indicate that organic molecules can be created without the mediation of enzymes,  Urey and Miller (1953). The transition is driven by  an external energy source ---  heat and electrical  discharge.

\ms
The theoretical basis for the transition is based on a model where the initial state is a random assembly of inorganic molecules. 
\ms
Evolutionary entropy, a measure of the cooperativity  of the interacting component, will be small.

\ms
We now assume that the system is furnished  a constant supply of energy, in the form of electric discharges. The effect of this constant source of energy is the generation of non--covalent interactions between  the inorganic molecules, and the emergence of components with differences in cooperativity, and consequently differences in evolutionary entropy. Competition between the local structures for the  energy source  entails that intermediates with increased evolutionary entropy,  and  concomitantly increased stability will emerge. The increase in evolutionary entropy will continue until a value is attained which exceeds the growth rate $r$. The critical level of organization in the model is described by a evolutionary  entropy which satisfies the condition
$$
H > r$$

\ms
\textbf{(II) Activated Biomolecules to Amphiphilic molecules} \textit{(Static Self--Assembly)}
\smallskip
The chemical reactions that define replication and metabolism take place in membrane--bounded compartments. The membranes consist of polymers, composed of organic components called amphiphiles.

\ms
Experimental studies indicate that lipid vesicles  self--assemble when a phospolipid is dispersed in aqueous  solution, Deamer (2011). The self--assembly of  membraneous vesicles is a spontaneous process. This ordering does not require a constant source of energy.  Self--organization in these systems is defined by static self--assembly. The equilibrium state is given by a configuration which maximizes evolutionary entropy. This is expressed by the condition:n
$$
\Delta H > 0
$$
\textbf{(III) Activated biomolecules to informational  polymers} (\textit{Dynamic self--assembly})

\smallskip
Modern biology is constructed from a large set of chemical reactions defined by replicating DNA, and catalytic proteins.

\ms
The complexity of  DNA and protein polymers suggest  that these systems are the derivatives of RNA polymers, the fundamental units of the RNA world. 

\ms 
The chemical bonds linking monomers into polymers are produced by the removal of water molecules. This chemical  reaction can only proceed with input from an external energy source.  Accordingly the transition from activated nucleotide monomers to RNA polymers can be described as dynamic self--assembly. Accordingly,  equilibrium configuration will satisfy the relation
$$
(H-r)\Delta H \ge 0
$$

\bigskip
\textbf{4.2 The transition from protocells to Cellular life}  \textit{(Entropic Principle of Evolution)}

\smallskip
A protocell consists of two key components: a protocell membrane that defines a spatially localized compartment, and an informational polymer consisting of RNA molecules capable of replication and catalysis. 

\ms
Modern cellular life is the integrated assembly of three classes of carbon--based polymers.
\begin{enumerate}
	\item[(i)] \textit{Amphiphilic molecules:} These are components, phospholipids, fatty acids, sterols, that contain covalently bonded components, with one part having high affinity for polar solvents, the other part, high affinity for non--polar solvents. These molecules are permeability barriers so that modern cells have complete control over the uptake of nutrients and the export of wastes.
\item[(ii)]\textit{ Proteins:} These carbon--based polymers are chains of amino acids. Proteins are characterized by their capacity to fold into three dimensional structures with catalytic activity.
\item[(iii)]\textit{ DNA:}  These polymers store and replicate genetic information. The major difference between DNA and RNA is the presence of a hydroxyl group in RNA, a property which makes RNA less stable.
\end{enumerate}

\ms
The transition from an RNA world, defined by RNA polymers, to cellular life, defined by the replacement of RNA by protein and DNA, can be explained in terms of  the \textit{Entropic Selection Principle.}

\ms
The transformations are delineated as follows

\smallskip
\textbf{(I)} \textit{RNA--based protocells:}

\smallskip
\begin{center}
	\includegraphics[width=2cm]{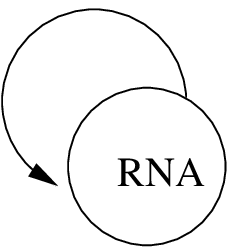}	
\end{center}

\ms
The RNA in  protocells function as unstable replicators and as inefficient catalysts. \\
The structure of RNA entails that the error rate of replication is high, a condition which impairs the 	effectiveness of RNA as an information carrier. 
The fragility of RNA entails that its stability is weak, a condition which impairs the catalytic effectiveness of the molecules. 

\ms
\textbf{(II)} The transition

\smallskip
\begin{center}
	\includegraphics[width=6cm]{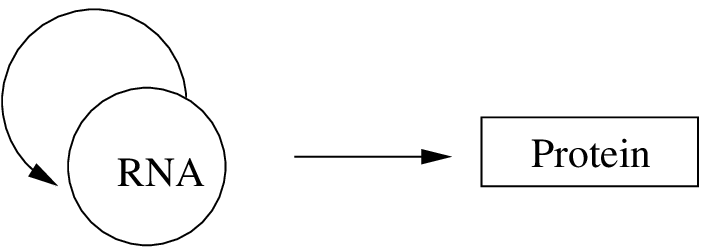}	
\end{center}		

\smallskip
Protein synthesis is assumed to have evolved from RNA--based protocells. The transition from RNA to  proteins confers a selective advantage on account of the superior catalytic activity of proteins.

\ms
\textbf{(III)} The transition

\smallskip	
\begin{center}
	\includegraphics[width=6cm]{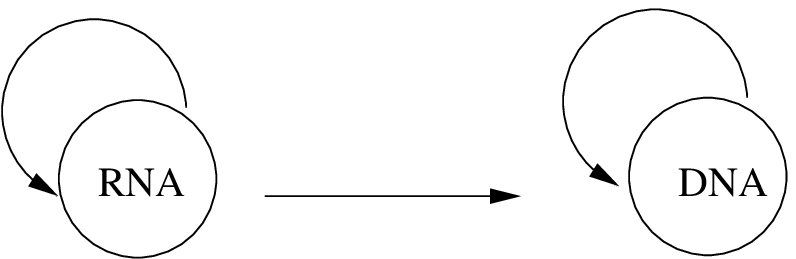}	
\end{center}		

\smallskip
Modern DNA replication occurs by a mechanism which involves the opening of the parental double helix into two separate strands so that each strand can serve as a template for the formation of a new strand.

\ms
The stability of DNA entails that the replacement of RNA by DNA  confers a selective advantage, as it ensures a low error rate and fidelity of the  information transfer process.

\ms
The transitions, RNA $\to$ Protein, and RNA $\to$ DNA, can be explained in terms of the\textit{ Entropic Selection Principle.} The selective advantage conferred by the transition  from RNA $\to$ Protein  is the enhancement  of catalytic activity. The selective advantage induced by the transition from RNA to DNA is the increase in the stability of the genome

\ms	
\begin{center}
	\includegraphics[width=15cm]{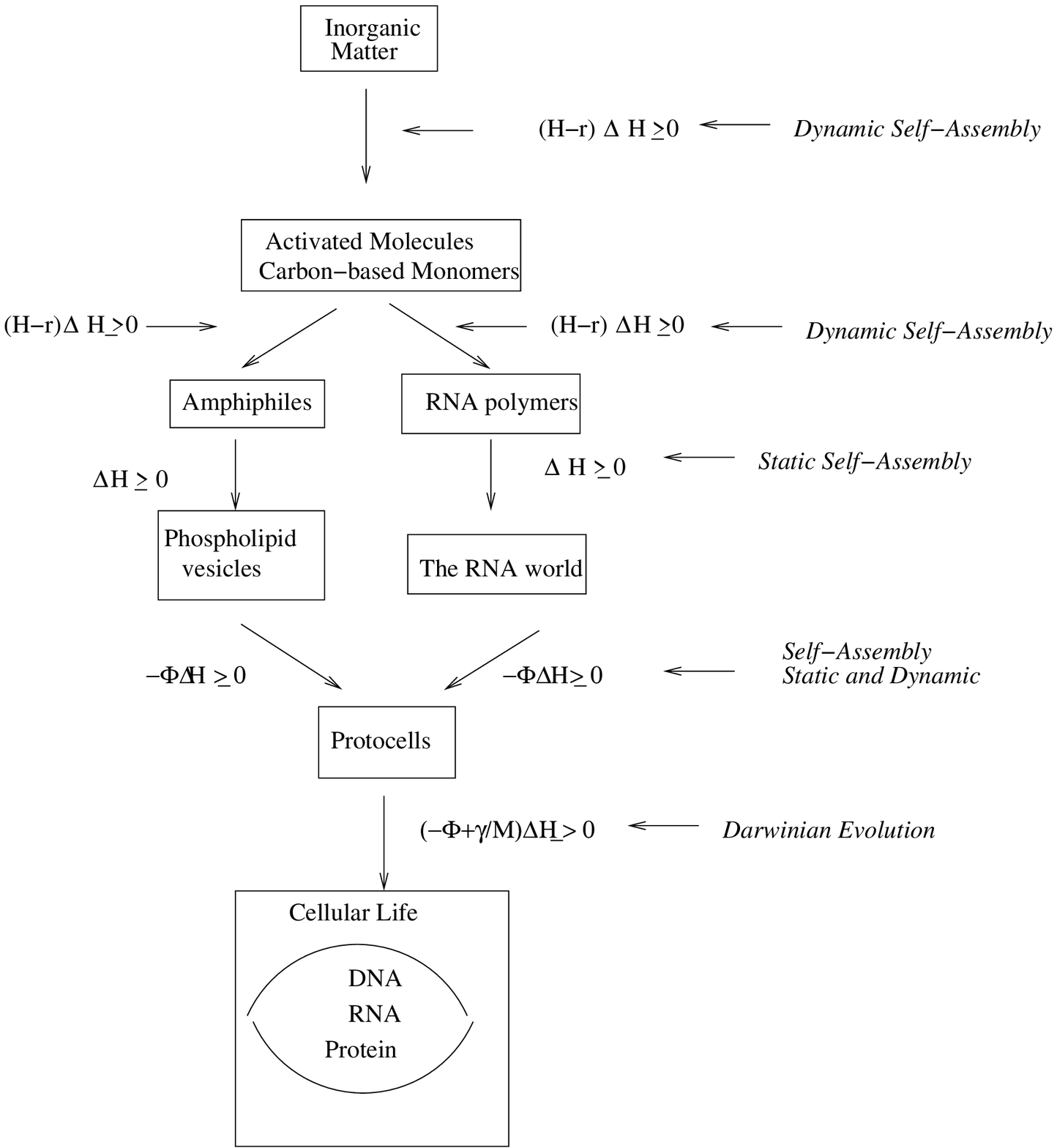}

\ms	
Fig. 3: The transition from inorganic matter to cellular life
\end{center}

\newpage
\centerline{\large{\bf 5. Conclusion}}

\ms
The fundamental microscopic units of inorganic matter are atoms and molecules. The collective behavior of these elements define the properties of solids, liquids and gases --- the entities that constitute condensed matter.

\ms
The \textit{Second Law of Thermodynamics} articulates the rule which describes  the collective behavior of aggregates of inorganic matter. This rule is epressed in terms of the measure of cooperativity,  Thermodynamic entropy.

\ms
The fundamental units of cellular life are a protocellular membrane that defines a spatially localized compartment, and the carbon--based polymers that allows for replication and metabolism. The collective behavior of protocells defines the adaptive properties of cells to the external  environment.

\ms
The \textit{Entropic Principle  of Evolution} enunciates  the rule which describes the collective behavior of aggregates of organic matter: activated macromolecules, cells, multicellular organisms. This rule is encrypted in terms of the statistical parameter, evolutionary entropy.

\ms
The physico--chemical pathways, as depicted in the transitions described in Sec. 4.,  modulate  the transition from inorganic matter to the chemical assemblies that define cellular life. These transitions derive their validity from the following analytical fact: \textit{The Entropic Principle of Evolution, which describes collective behavior of organic matter,  is a generalization of the Second Law, which pertains to collective behavior in inorganic matter.}

\ms
We will elucidate the transition from inorganic matter to cellular life by providing a qualitative account of the relation between the Entropic Principle of Evolution,  and the Second Law of Thermodynamics.

\bigskip
\textbf{5.1 Collective behavior: Inorganic matter}

\smallskip
A characteristic property of the collective behavior of aggregates of inorganic matter is the irreversible flow of thermal energy in isolated systems. 

\ms
Clausius made a fundamental contribution to the elucidation of the phenomenon by showing that thermal  irreversibility entails the existence of a property of matter which he called entropy.

\ms
The Clausius entropy, denoted $S_C$, is analytically expressed by the relation
$$
dS_C = \frac{dQ}{T}
$$
$dS_C$ denote the change in entropy, $T$ is the absolute temperature and $Q$ has units of energy. The relation $dQ > 0$ means that heat flows into the system; $dQ < 0$ means that heat flows out of the system. Consequently,  $dS_C$, the change in entropy will be positive or negative when heat flows into or out of the system.

\ms
The problem of determining a statistical mechanics  explanation of this irreversible behavior was resolved by Boltzmann. This resolution was based on a model which assigns to each microstate of a macroscopic system --  a solid, liquid or gas, a number $S_B$, the Boltzmann entropy given by
$$
S_B=k\log W
$$
The quantity $W$ denote the total number of microstates of a system consistent with a given macrostate. 

\ms
If we assume that the particles in the macroscopic aggregate occupy small volumetric cells, $i=1,2,\ldots s$ of phase space, with occupation numbers $n_i$, and that the total number $N$ of particles is conserved, then
$$
N=\sum^s_{i=1} n_i
$$
The quantity $W$, which denote the nmber of instantaneous microstates is given by
$$
W=\frac{N!}{n_1!n_2!\ldots n_s}
$$
Boltzmann's analysis of changes in the thermodynamic  entropy  assumed that the individual particles were subject to the laws of  Newtonian mechanics. The analysis showed that as the configuration changes, due to the collision of the particles that comprise the macroscopic system, the entropy $S_B$ increases. We write
$$
\Delta S_B > 0
$$
Boltzmann furthermore showed that there is a direct connection between the microscopically defined thermodynamic entropy $S_B$, and the phenomenologically defined entropy of Clausius.

\ms
For large $N$, both quantities coincide. This identification of the two measures of entropy explains the observation embodied in the Second Law of Thermodynamics, namely:  when a constraint is lifted from a system which is isolated in terms of energy and matter, the system evolves towards a state of maximumg entropy.

\bigskip
\textbf{(5.2) Collective behavior: Organic Matter}

\smallskip
Aggregates of organic matter --- activated macromolecules, cells, and higher organisms, differ from inorganic matter in terms of the cooperativity of the components, and their capacity to evolve by increasing or decreasing this cooperativity.

\ms
Collective behavior in population of cells, a unit defined by replicative and metabolic entities, is determined by following processes: 
\begin{enumerate}
	\item \textit{Variation:} Random changes in the informational molecules that encode the behavior of the individual cells
\item \textit{Selection:} Competition between the variant and incumbent cells  for the energy provided by the resource which define the external environment
\item \textit{Inheritance:}	The correlation between the behavioral properties of ancestors and descendants  
\end{enumerate} 
The parameter $\Phi$ is given by the identity
$$
\Phi = r-H
$$

Cooperativity in the population  is described by the statistical parameter, evolutionary entropy, $H$. This parameter is given by $H=S/T$ where
$$
S=-\sum_{\tilde{\alpha} \in C_\alpha}p_{\tilde{\alpha}}\log p_{\tilde{\alpha}}; \, \, 
T=\sum_{\tilde{\alpha} \in C_\alpha}|\tilde{\alpha}| p_{\tilde{\alpha}}
$$
The changes in $H$, as one population replaces another by mutation and natural selection is described by the following rule
$$
(-\Phi+\gamma/M)\Delta H \ge 0\leqno(8)
$$
The statistical parameters, $\Phi$ and $\gamma$, are correlated with the external energy resources. The parameter $M$ denote the population size.

\bigskip
\textbf{5.3 The Second Law of Thermodynamics and the Entropic Principle  of Evolution}

\smallskip
The Second Law of Thermodynamics asserts that in isolated systems, the thermodynamics entropy,  the statistical parameter $S_B$, increases.

\ms
The Entropic Principle of Evolution  asserts that  the change in evolutionary entropy  is contingent on the resource endowment, and the population size, and predicted by  extremal states of evolutionary entropy.

\ms
The statistical parameters $\Phi$ and $\gamma$ are related to the resources generated by the external energy source. The parameter $\Phi$, the resource production rate, is given by the identity
$$
\Phi=r-H
$$
The parameter $\gamma$ describes the temporal correlation of the resource production rate.

\ms
We  elucidate the relation between the two principles by imposing various constraints on the statistical parameters that determine the evolution of collective behavior in organic matter.

\ms
\textbf{(i)} \textit{The condition $\gamma=0$}\\
This constraint entails that the forces which effect cooperation between the components that define the aggregate are reversible. Collective behavior in this system is self--organizing. The relation given by (8) reduces to the expression
$$
-\Phi\Delta H > 0\leqno(9)
$$
\textbf{(ii)} \textit{The condition $r \neq 0$}\\
The constraint  entails that the energy is dissipative. The expression (9) now reduces to the relation
$$
(H-r)\Delta H \ge 0\leqno(10)
$$
The relation (10) describes \textit{dynamic} self--assembly 

\ms
\textbf{(iii)} \textit{The condition $r = 0$}\\
The relation in this case reduces to the expression
$$
\Delta H \ge 0\leqno(11)
$$
This relation describes \textit{static} self--assembly. The equilibrium state is characterized by the state which maximizes evolutionary entropy.

\ms
\textbf{(iv)} \textit{The condition $\tilde{N} \to \infty$}\\
Here $\tilde{N}$ denote the number of microstates. We infer from the condition that the evolutionary entropy $H$, and the thermodynamic entropy $S_B$ coincide.

\ms
The relation (11) now reduces to the expression
$$
\Delta S_B \ge 0\leqno(12)
$$
This argument indicates that the Entropic Principle of Evolution  is a generalization of the Second Law.

\ms
The analytic relation between the Entropic Principle of Evolution and the Second Law of Thermodynamics constitutes the rationale for the model which describes the transition from inorganic matter to cellular life.

\newpage
\centerline{\large\bf  References}
\begin{enumerate}
	\item Alberts, B.M. (1986): The function of the Hereditary materials: Biological Catalyses reflect the cellss' evolutionary history.\textit{ Amer. Zool.} 26, 781--796.
	\item Andrien D. and Gaspard, P. (2008): Non--equilibrium generation of information in copolymerization processes.\textit{ Proc. Nat. Acad. Sciences}, 105, 9516--9521.
	%\item Anfinsen, C.B. (1973): Principles that govern the folding of protein chains. \textit{Science} 181, 223--230.
	%\item Arnold, L., Demetrius, L. and M. Gundlach (1994): Evolutionary Formalism for products of positive random matrices. \textit{Ann. Appl. Probab.} 4, 859--901.
	%\item Baker, D. (2000): A surprising simplicity to protein folding. \textit{Nature}, 405, 39--42.
	%\item Bonabeau, E., Dorigo, M. and G. Therauly (1999): Swarm intelligence from natural to artificial systems. OUP. ISA
	%\item Ben--Naim, A. (2020): Entropy and Time. \textit{Entropy} 22, 430
	\item Boltzmann, L. (1896): Vorlesungen \"uber Gas Theorie. Barth. Leipzig, Germany.
	%\item Bowen, R. (1975): Equilibrium States and the Ergodic Theroy of Anosov diffeomorphism. Springer Verlag.
	%\item Camazine, S. (2003): Self--Organization in Biological Systems. Princeton Studies in complexity. Princeton University Press.
	\item Cech, T.R. (1985): Self--splicing RNA: Implications for evolution. \textit{Int. Rev. Cytol.} 93, 3--22.
	\item Chen, I., Roberts, R.W. and J.W. Szostak (2004): the emergence of Competition between Model Protocells. \textit{Science} 305, 1474--1476.
	\item Clausius, R. (1867): Mechanical Theory of Heat. John van Voorts, London.
	\item Copley, S., Smith, E. and H.J. Morowitz (2007): The origin of the RNA world. Co--evolution of genes and metabolism. \textit{Bioorganic Chemistry} 35, 430--443.
	%\item Crick, F. (1981): Life itself: its nature and origin. Simon and Schuster.
	\item Crooks, GE. (1999): Entropy Production Fluctuation. Theorem and the Nonequilibrium Work Relation for Free Energy Differences: \textit{Physical Review} E-60  2721.
	\item Damer, B. and D. Deamer (2020): The Hot Spring Hypothesis for an Origin of Life. Astrobiology 20, No. 4, 429--454.
	\item Davies, P. (1998):  The origin of Life. London:  Penguin Books.
	%\item Davies, P. (1998): The fifth miracle: The Search for the origin of Life. London:  Penguin Books.
	\item Davies, P., Rieper, E. and J. Tuszynski (2012): Self--Organization and Entropy reduction in a living cell. \textit{Biosystems }Vol. 111, 1--10.
	\item Deamer, D.W. (1997): The first Living systems: a bioenergetic perspective. \textit{Microbiology and Molecular Biology Reviews,} Vol. 61, 231--261.
	\item Deamer, D. (2011): First Life; University of California Press.
	\item Deamer, D. and  Szostak; H. (2010): The origins of Life; Cold Spring Harbor Press, Cold Spring Harbor.
	\item de Duve, C. (1995): Vital Dust: Life as a Cosmic Imperative. Basic Books.
	\item Demetrius, L. (1983): Statistical mechanics and population Biology. \textit{Jour. Stat. Phys.} Vol. 30, 709--753. 
	\item Demetrius, L. (1974):	Demographic parameters and natural selection. \textit{Proc. Natl. Acad. Sci.} 4645--4649.
	\item Demetrius, L. (1975): Natural selection and age--structured population.\textit{ Genetics} 79, 533--544.
	\item Demetrius, L. (1984):Self--Organization in macromolecular systems: The notion of adaptive value. \textit{Proc. Natl. Acad. Sci. USA} 81, 6068--6072
	\item Demetrius, L. (1997):	Directionality Principle in Thermo--dynamics and Evolution. \textit{Proc. Natl. Acad. Sci.} Vol. 94, 3491--3498.
	%	\item Demetrius, L. and T. Manke  (2005): Robustness and Network Evolution \textit{Physica. A.} 682--696.	
	\item Demetrius, L.; Gundlach, V.M. and G. Ochs (2009): Invasion exponents in biological networks.\textit{ Physica A.} 388--572.
	\item Demetrius, L. (2013): Boltzmann, Darwin and Directionality Theory. \textit{Physics Reports}, Vol. 530, 1--85.
	%\item Demetrius, L. and M. Gundlach (2014): Directionality Theory and the Entropic Selection Principle.\textit{ Entropy} 15, 5428--5622.
		%\item Demetrius, L. and S. Legendre (2013): Evolutionary entropy predicts the outcome of selection: Competition for resources that vary in abundance and disposition.\textit{ Theor. Popul. Biology} 83, 39--54.
	\item Demetrius, L.	 and C. Wolf (2022): Directionality Theory and the Second Law of Thermodynamics, \textit{Physica A., Statistical Mechanics and its Applications. }Vol 598, 127325.
	\item Demongeot, J. and H. Seligmann (2022): Evolution of small and large ribosomal RNAs from accretion of tRNA subelements.\textit{ BioSystems} Vol. 222.104796.
	\item Demongeot, J. and M. Thellier (2023): Primitive Oligomeric RNAs at  the Origins of Life on Earth. \textit{Int. Jour. Mol. Sci.} Vol. 24, 2274.
	\item Dyson, F. (2004): Origin of Life, Cambridge University Press.
	\item Eigen, M. (1992): Steps towards Life: A perspective on  Evolution. Oxford University Press.
	\item Endres, R.G. (2017): Entropy production selects non--equililbrium states in multistable systems. \textit{Scientific Reports, }Vol 7, 14437.
	\item England, J.L. (2013): Statistical Physics of Self--Replication. \textit{Jour. Chem. Physics }139, 121923.
	%\item Fisher, R.A. (1930): The Genetical Theoryof natural selection, Dover, New York.
	%\item Haken, H. (1977): Non--equilibrium phase transitions and self--organization in physics, chemistry, and biology in Synergetics: An Introduction. Springer, Berlin.
	%\item Hannezo, E. et al. (2017): A unifying Theory of Branching Morphogenesis. \textit{Cell }171, 242--255.
	\item Hartman, H. (1998): Photosynthesis and the Origin of Life. \textit{Origins of Life and the Evolution of the Biosphere,} Vol. 28, 519--521.
	\item Higgs, P.G. and N. Lehman (2015): the RA World: molecular cooperation at the origins of life. \textit{Nature Reviews Genetics} Vol. 16, 7--17.
	\item Jarzynski, C. (2011): Equalities and Inequalities: Irreversibility and the Second Law of Thermodynamics. \textit{Ann. Rev. Condens. Matter Phys.} 2: 329--351.
	%\item Johnston, I.G. et al. (2022) Symmetry and simplicity spontaneously emergence from the algorithmic nature of evolution.\textit{ Proc. Natl. Acad.} Vol. 119,No. 11	
	%\item Karsenti, E. (2008): Self--organization in cell Biology: a brief history. \textit{Nat. Rev. Mol. Cell. Biol.} 9, 255--262.
	\item Kauffman, S. (1993): The Origins of Order: Self--Organization and Selection in Evolution. Oxford University Press.
	%\item	Kirschner, M.  and T. Mitchison (1986): Beyond Self Assembly: From Microtubules to Morphogenesis. \textit{Cell}, Vol 45, 329-342.
	\item Kornberg, A., Bartsch, L., Jackson, J. and H.G. Khorona (1964): Enzymatic synthesis of Deoxyribonucleic acid, XUI. Oligonucleotides as templates and the mechanism of replication.\textit{ Biochemistry}, 51, 315--323.
	\item Lazcano, A. and S.L. Miller (1996): The Origin and early evolution of Life: Prebiotic Chemistry, the Pre--RNA World, and Time.\textit{ Cell} 85, 793--798.
	\item Lehn, J.M. (2012): Constitutional Dynamic Chemistry: Bridge from Supramolecular Chemistry to Adaptive Chemistry. \textit{Top. Curr. Chem.} 322, 1--32.
	%\item Li, Y, and M.E. Cates (2020): Steady state entropy production rate for scalar Langevin field theories. \textit{Journal  Statistical Mechanics. }
	%\item Meinhardt, H. (1982): Models of Biological Pattern. Formation London, Academic Press.
	\item Michaelian, K.l (2011): Entropy Production and the origin of Life. \textit{Journal of Modern Physics}, Vol 2, 595--601.
	%\item Misteli, T. (2001): The concept of self--organization in cellular architecture. \textit{Journ. Cell Biology,} Vol 55, 181--185.	
	\item Morowitz, H. and E. Smith (2007): Energy flow and the Organization of Life.\textit{ Complexity }13, 51--59.
	\item Nicolis, G. and I. Prigogine (1977): Self--organization in Non--equilibrium systems: From Dissipative Structures to Order through fluctuation. Wiley.
	\item Onsager, L. (1931): Reciprocal relations in irreversible processes. \textit{Phys. Rev.} 37,\textit{} 405--426.
	%\item Ornes, S. (2017): How nonequilibrium thermodynamics speas to the mystery of life. \textit{Proc. Natl. Acad. Sci.} Vol 114, No. 3, 423--424.
	\item Prigogine, I (1969): Structure, dissipation and Life. In: Theoretical Physics and Biology ed. M. Marois, Amsterdam.
	\item Pross, A. (2012): What is Life? How chemistry become Biology. Oxford University Press.	
%	\item Ruelle, D. (1978): Thermodynamic Formalism, Addison--Wesley Reading MA.
	%\item Saha, T. and M. Galic (2018): Self--Organization across scales: from molecules to organisms.\textit{ Phil. Trans. R. Soc. B.} Vol. 373: 20120118
	\item Schneider, E.D. and J.J. Kay (1994): Life as a manisfestation of the Second Law of Thermodynamics. \textit{Math. Comput. Model} 19, 25--48.
	%\item Shaknovich, E. (1994): Theoretical studies of protein folding thermodynamics and kinetics. \textit{Current Opinion in Natural biology}, Vol. 7, 27--40. 
	\item Szostak, J., Bartel, D.P. and P. Luigi Luisi (2001): Synthesizing Life.\textit{ Nature}, Vol. 409, 387--390.	
	%\item Schr\"odinger, E. (1967): What is Life. Cambridge University Press.
	%\item Tabony, J., Glade, N. and J. Demongeot (2022): Microtubule self--organization: a biological example of emergent phenomena in a complex system.\textit{ Recent. Res. Devel. Biosphys. Chem.} 3, 11--53.
	%\item Tabony, J., Glade, N. and Papasait, C. and J. Demongeot (2004): Microtubule self--organization as an example of the development of order in living systems.\textit{ J. Biol. Phys. Chem.} 4, 50--63.
	%\item Turing, A.M. (1952): The chemical basis of morphogenesis. \textit{Phil. Trans. Royal Socciety } 237, 37--52.
	\item H. Urey and S. Miller (1953): Production of Amino acids under possible primitive earth conditions. \textit{Science,} Vol. 117, 528--529.
	\item Walker, S.I. (2017): Origins of Life: a problem for Physics, a key issues review. \textit{Review Prog. Phys.} 80--09260180(9).
	%\item Whitesides, G. and B. Grzybowski (2002): Self--assembly at all Scales. SCIENCE, Vol 295, 2418--2423	
\end{enumerate}
\end{document}